# Improved data analysis of the internal background measurements of $^{40}$Ca$^{100}$MoO$_4$ scintillation crystals

**Nikita Khanbekov**[a,*], **Vladimir Alenkov**[b], **Aleksander Burenkov**[a], **Oleg Buzanov**[b], **HongJoo Kim**[c], **Vasily Kornoukhov**[a], **JungHo So**[c]

[a] Institute for Theoretical and Experimental Physics
ul. B. Cheremushkinskaya 25, Moscow, 117218 Russia
[b] JSC FOMOS-Materials
ul. Buzheninova 16, Moscow, 107023 Russia
[c] Physics Department, Kyungpook National University
Daegu 702-701, Republic of Korea

E-mail: xanbekov@gmail.com

ABSTRACT: The sensitivity of neutrinoless double beta ($0\nu2\beta$) decay experiments is mainly depended on the internal background of a detector which, in its turn, is defined by the purity of material and possibility for selection of background events. The AMoRE (Advanced Mo based Rare process Experiment) collaboration plans to use $^{40}$Ca$^{100}$MoO$_4$ scintillation crystals as a detector for search of $0\nu2\beta$ decay of $^{100}$Mo isotope. A purpose of this paper is further investigation of internal background of $^{40}$Ca$^{100}$MoO$_4$ scintillation elements with a low background setup at YangYang underground laboratory. We present new approaches for selection of background events from analyzing data and latest results of measurements of intrinsic background of $^{40}$Ca$^{100}$MoO$_4$ crystals as an example of new technique application.

---

[*] Corresponding author.

# Contents

## 1. Introduction

Discovery of the neutrino oscillation means that neutrinos have non-zero mass [1]. It's become a new impetus for further searches of physics beyond the Standard Model. Detection of 0ν2β decay would give an opportunity to determine the effective mass of neutrino <m$_ν$> and would confirm that neutrino is a Majorana particle, i.e. neutrino is identical anti-neutrino.

One part of Heidelberg-Moscow collaboration claimed to observe 0ν2β decay of $^{76}$Ge isotope [2]. Nowadays the GERDA experiment sets a goal to confirm or disprove this result [3]. Several groups are carrying on experiments with other isotopes. EXO-200 [4] and KamLand-Zen [5] experiments search for the 0ν2β decay of $^{136}$Xe isotope.

Next generation 0ν2β experiments need the high energy resolution, low background of a detector and many tens or hundreds kilograms of the working isotope with the high transition energy $Q_{ββ}$. $^{100}$Mo isotope is one of the best candidates for 0ν2β experiments because of the highest value of $Q_{ββ}$ = 3034 keV and possibility for production this isotope in a big amount by centrifugation method [6], [7].

Appropriate detector for the experiment with $^{100}$Mo isotope which was chosen by AMORE collaboration [8] is a scintillation bolometer based on $^{40}$Ca$^{100}$MoO$_4$ single crystals. The fractional mass of the molybdenum element in the crystal is relatively high (about 50%). The maximum energy of γ-background from natural long-lived isotopes is 2615 keV ($^{208}$Tl from $^{232}$Th-chain). Thus, on the condition that the resolution of a detector is good, external γ-background doesn't contribute to the ROI (the Range of Interest) for $^{100}$Mo.

On the contrary the two-neutrino double beta decay of $^{48}$Ca ($Q_{ββ}$ = 4271 keV) will give unavoidable background, limiting the sensitivity of the experiment with enriched $^{40}$Ca$^{100}$MoO$_4$. Since natural calcium contains 0.187 % of $^{48}$Ca we need to use calcium depleted on $^{48}$Ca isotope [6]. Another problem is the internal background due to the presence of dangerous radionuclides inside the detector. There are two isotopes with decay energy higher than 3034 keV: $^{208}$Tl ($Q_β$ = 5001 keV) from $^{232}$Th –chain and $^{214}$Bi ($Q_β$ = 3272 keV) from $^{238}$U-chain. Two approaches are used to reduce this background: purification of material to remove radioactive impurities and selection of background events by sophisticated data analysis.

Previously the AMORE collaboration, in order to estimate the radioactive contamination, used a time-amplitude analysis [9], which exploits the energies and time difference between primary and secondary signals to select specific fast sequences of decays in the U/ Th chains. Thus we selected the decay events of $^{214}$Po, $^{216}$Po, and $^{220}$Rn isotopes. This study reports a new technique (and its application) of data analysis of scintillation signals, which allows us to solve a problem of pile-up events from the $^{212}$Bi- $^{212}$Po and $^{214}$Bi- $^{214}$Po decays [9], [10].

## 2. Production of $^{40}$Ca$^{100}$MoO$_4$ crystals and scintillation elements

### 2.1 Production of $^{100}$Mo and $^{48}$Ca isotopes.

Enriched molybdenum (with 96.1 % of enrichment by $^{100}$Mo isotopes) is produced by the JSC Production Association Electrochemical plant (Zelenogorsk, Krasnoyarsk region, Russia [11]) and supplied in the form of molybdenum oxide $^{100}$MoO$_3$. The results of ICP-MS measurements show that the enriched material is very pure. Concentrations of $^{238}$U and $^{232}$Th in the oxide do not exceed 0.07 and 0.1 ppb, respectively.

Calcium carbonate $^{40}$CaCO$_3$ enriched of $^{40}$Ca (99.964 %) and depleted of the $^{48}$Ca (content is ≤ 0.001 %) is produced by FSUE Electrochimpribor (Lesnoy, Sverdlovsk region, Russia [12]). The concentration of $^{238}$U and $^{232}$Th in the enriched powder measured by ICP-MS is below 0.2 and 0.8 ppb, respectively. However, HPGe measurements at Baksan Neutrino Observatory showed that the activity of $^{226}$Ra and its daughter isotopes in the decay chain was on a level of hundreds mBq/kg [13]. For that reason the raw materials were subjected to additional purification. A new technique of purification of calcium carbonate in the form of calcium formate Ca(HCOO)$_2$ [14] allowed to reduce a content of $^{40}$K, $^{208}$Tl, $^{228}$Ac, $^{226}$Ra ($^{214}$Bi) in 20, 8, 160 and 5 respectively in compare with the standard procedure of purification on Electrochimpribor plant.

### 2.2 Production of $^{40}$Ca$^{100}$MoO$_4$ crystals.

$^{40}$Ca$^{100}$MoO$_4$ crystals were grown by Czochralski technique by JSC FOMOS-MATERIALS [15]. The specific process, which calls recrystallization, consisted of two steps. On the first step two single crystals (boules) were pulling out of mixture in platinum crucible. Then, on the second step, the grown boules were using as a raw material for pulling of a final single crystal. The recrystallization process allows to reduce radioactive contamination in the crystal [14]. Thus $^{226}$Ra content was decreased from 260 mBq/kg in initial calcium carbonate $^{40}$CaCO$_3$ to 6.4 mBq/kg in a $^{40}$Ca$^{100}$MoO$_4$ crystal (the reduction factor of 40 times). Crystals were bluish just after growing. To decolourize crystals they were annealed in oxidative (oxygen) atmosphere [6], [14].

Cone parts of crystals were cut and then their ends were optically polished to produce scintillation elements. Thus, studied scintillation elements have cylindrical shape with elliptic base, unpolished side surface and optical polished butt-ends. In the future $^{40}$Ca$^{100}$MoO$_4$ single crystals, which produced by this technique from more pure raw materials, will be used for production of scintillation elements of cryogenic scintillation detector of AMORE Collaboration.

Two $^{40}$Ca$^{100}$MoO$_4$ scintillation elements were used for our measurements: SB29 and S35. The SB29 crystal was produced with recrystallization process, but S35 were grown directly from $^{40}$Ca$^{100}$MoO$_4$ charge (raw material). Dimensions and mass of the crystals are presented on the Table 1.

Table 1. Properties of $^{40}$Ca$^{100}$MoO$_4$ scintillation elements.

|  | SB29 | S35 |
|---|---|---|
| Height, mm | 48/51 | 42 |
| D big, mm | 50 | 44 |
| D small, mm | 42.5 | 39.5 |
| Mass, g | 390.35 | 259 |
| Annealing, h | 240 | 48 |
| Color | bluish | colorless |

## 3. Low-background experimental setup

Background measurements of $^{40}Ca^{100}MoO_4$ scintillation elements (below in the text we will use word "crystals") were carried out in YangYang underground laboratory (Y2L) in depth of 700 m (approximately 2000 m of water equivalent). Information about laboratory radioactive background was given in [16], [17]. In order to decrease external background the $4\pi$ gamma veto system was used [18], [19]. The system consisted of 14 CsI(Tl) crystals (12 long crystals for sides and 2 short crystals for front-end caps) which were able to register the internal and external gammas. The veto-system rejected the coincidence events from $^{40}Ca^{100}MoO_4$ and one or several CsI(Tl) crystals. The system was shielded by lead (10 cm thick, Fig. 1) During the measurements, the whole internal cavity of system was continuously flushing by gas (at rate of 4 l/min) to prevent the contamination by $^{222}Rn$ present in air.

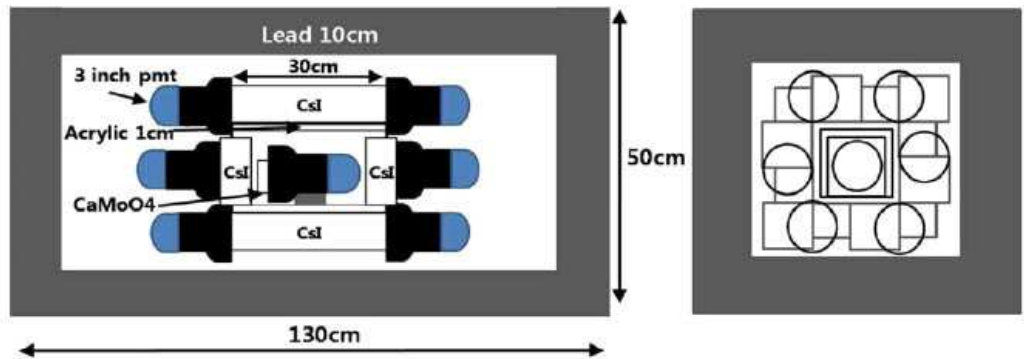

Fig. 1. Scheme of the setup for background measurements of $^{40}Ca^{100}MoO_4$ crystals [19].

The scintillation light from $^{40}Ca^{100}MoO_4$ crystals were collected by 7.62 cm diameter green enhanced PMT (Electron tube Ltd.). Crystals were wrapped by Teflon tape and connected to PMT using optical grease. Signals from $^{40}Ca^{100}MoO_4$ and from CsI(Tl) veto-system were amplified 10 times and digitized by 400 MHz FADC and 64 MHz FADC respectively. The recorded time-window from $^{40}Ca^{100}MoO_4$ was 82 µs (the decay constant of CaMoO4 is 16.4 µs). Veto-system data were recorded simultaneously with $^{40}Ca^{100}MoO_4$ signals [9].

## 4. Selection procedure for SB29 and S35 scintillation elements

**4.1.** We used background data which was collected during 50 days for S35 crystal and 90 days for SB29 crystal. First of all we worked out $^{214}Bi$ and $^{208}Tl$ ($^{220}Rn$) problem by time-amplitude analysis [6], [10]. In order to select corresponding events from raw background data, we used the following decay chains: $^{214}Bi$ ($Q_\beta$ = 3.27 MeV, $T_{1/2}$ = 19.9 min) → $^{214}Po$ ($Q_\alpha$ = 7.83 MeV, $T_{1/2}$ = 164 µs) → $^{210}Pb$ and $^{220}Rn$ ($Q_\alpha$ = 6.41 MeV, $T_{1/2}$ = 55.6 s) → $^{216}Po$ ($Q_\alpha$ = 6.91 MeV, $T_{1/2}$ = 145 ms) → $^{212}Pb$. To find $^{214}Bi$ events we checked signals in the time window after $^{214}Bi$ event which equal to $5xT_{1/2} \approx 0.8$ ms (there $T_{1/2}$ stands for the half life of $^{214}Po$ isotope). If the event with the energy release which corresponds to $^{214}Po$ in the electron equivalent energy is found we conclude that previous event was from $^{214}Bi$ decay. Thus we are able find up to 85 % of $^{214}Bi$ decays [6]. Last results of measurement of $^{214}Bi$ (1.74 mBq/kg) and $^{208}Tl$ (0.26 mBq/kg) content for S35 crystal were reported in [9]. However that time we were not able to count pile-up signals from two consecutive events in the same 82 µs window.

Two approaches were developed to solve this problem. The first one is based on possibility to select normal events by waveform analysis. If we make partial charge integration within a small time window inside the signal, for signals without overlapping it must be a roughly linear dependence of the energy of the whole signal from value of the integration in the new time window inside the signal. But pile-up signals fall out of this dependence (Fig. 2). This fact gives us an instrument for selection of pile-up events.

Using this method we identified as pileups 60.5 % of background events in the ROI for 0ν2β-decay (2935-keV < E < 3135-keV, taking into account of the energy resolution of S35 crystal).

In the next section we will discuss the second approach for identification of pileup events.

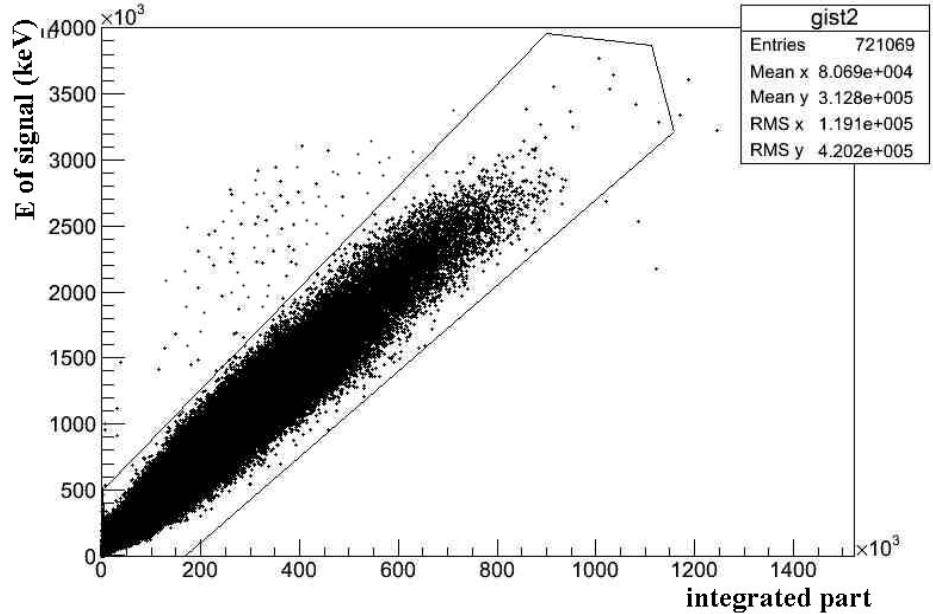

Fig. 2. Plot of energy of a signal (keV) versus energy of the selected part inside the signal.

**4.2** As it was already shown scintillation light from $^{40}Ca^{100}MoO_4$ crystals consist of several light-emission components with different decay constants. Shapes of these components and values of decay constants are different for α-events and γ(β)-events [6]. On the other hand an average decay time can be used for selection of pileup events in the data.

The average decay time of scintillation signal was calculated as

$$< t > = \sum E_n * n / E$$

where $E_n$ is the energy in one bin in the signal histogram, **n** is a number of bins from the beginning of the signal and **E** is the total energy of the signal. Dependence of the total energy versus average decay time for S35 crystal is shown on Fig. 3. Colour gradient is determined by the density of points in the plot region (by ROOT analysis). Elliptical circles indicate the events from α-decay of $^{210}$Po and $^{214}$Po isotopes. As we can see centers of these concentrations are shifted left on the center of a main distribution (which consists from gamma events).

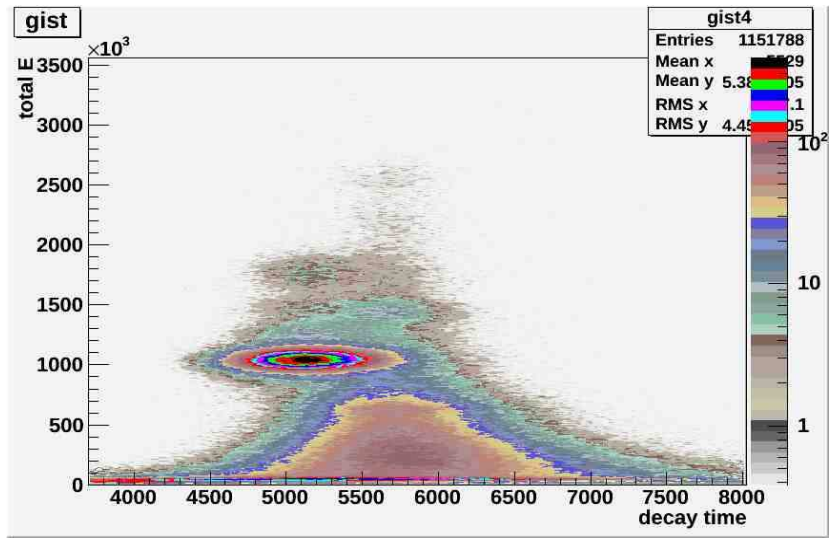

Fig. 3. Dependence of the total energy versus average decay time for S35 crystal. Colour gradient is determined by the point density on the given region of the plot.

Pileup signals must have longer decay time and are located more to the right from the mean β(γ)-time (Fig. 4). It's worth to notice that most of events on the right part of the picture are above the level of energy of $^{214}$Po events (1.93 MeV). It also indicates that the found events are pileups. Using standard ROOT function we fitted the edge of the distribution on the plot by hyperbola in order to select pileup events.

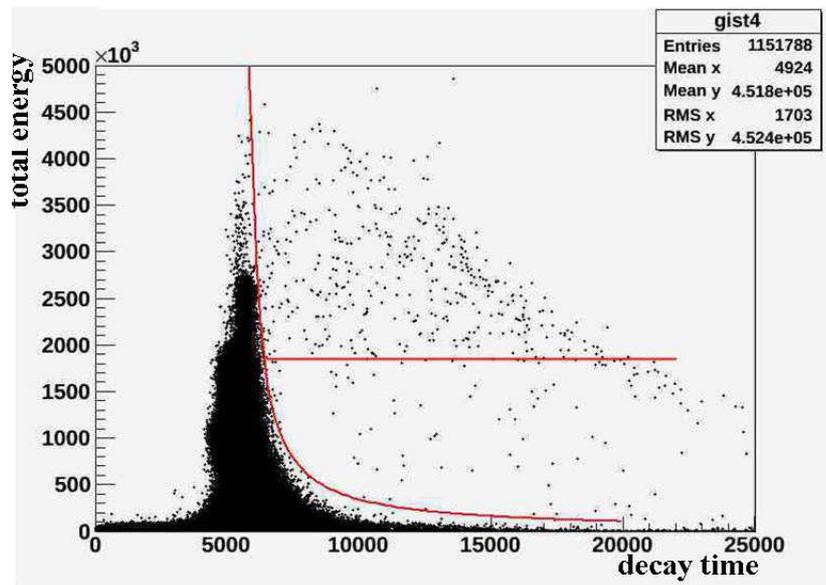

(a)

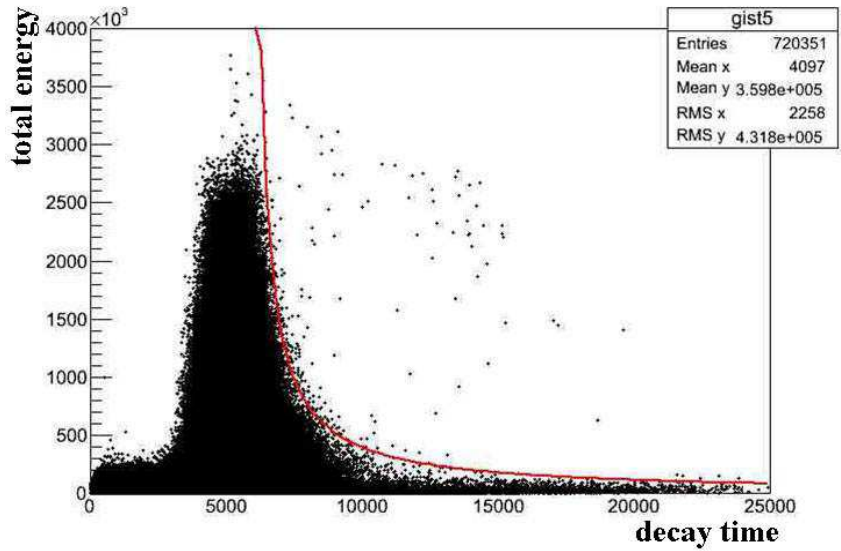

(b)

Fig. 4. Dependence of the total energy versus the average decay time for S35 (a) and SB29 (b) crystals with hyperbolic fitting.

Using this method we selected 64.2 % events from the ROI of S35 crystal. Due to small internal background of SB29 there was no difference for results of the both approaches for this crystal.

## 5. The background index of SB29 scintillation element

The measured FWHM energy resolution of SB29 is 26.5% with 662 keV γ-rays ($^{137}$Cs source) at room temperature. An energy spectrum is presented on Fig. 5.

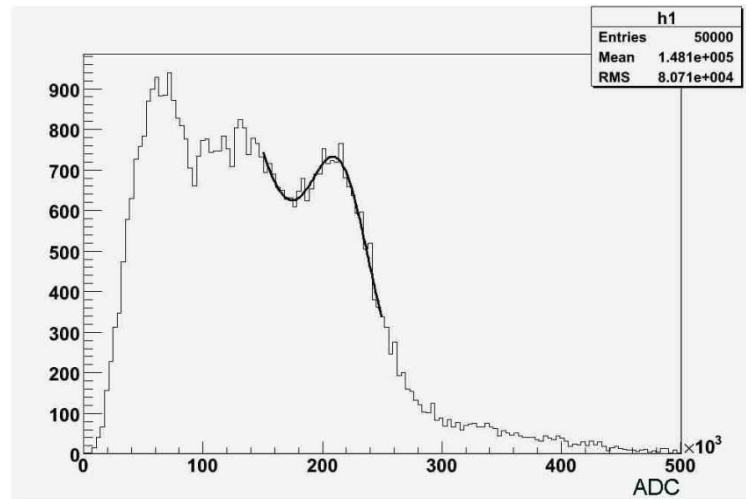

Fig. 5. Energy spectrum ($^{137}$Cs source) measured with SB29 crystal.

The background spectrum of SB29 crystal is presented on Fig. 6.

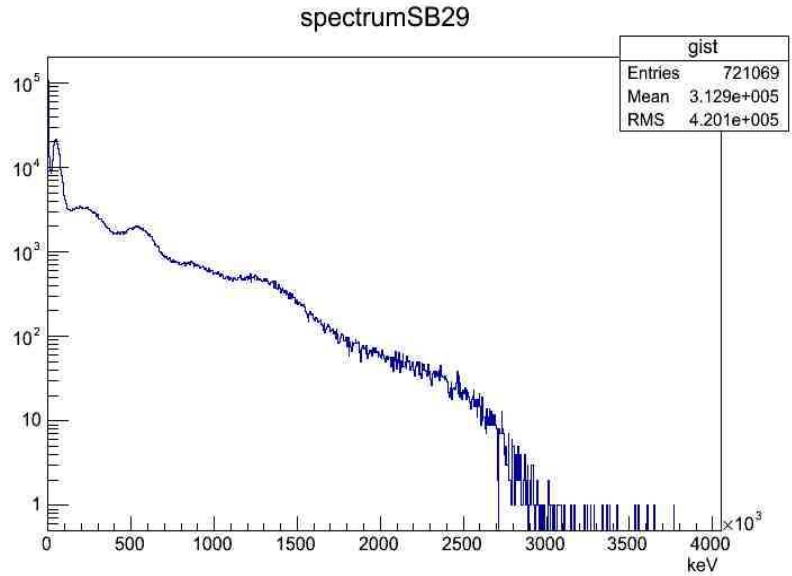

Fig. 6. Background spectrum of the SB29 crystal after 90 days of measurements.

In spite of worse energy resolution SB29 crystal is an example of $^{40}Ca^{100}MoO_4$ crystal with high purity and, as a consequence, low internal background. After 90 days of measurements we registered only 21 events in the ROI (2847-keV < E < 3223-keV) and 3 events among them were identified as pileups events by methods we described in section 4 (Fig. 7). Thus, the background index was defined equals 0.5 kev$^{-1}$ kg$^{-1}$ year$^{-1}$, which gives the sensitivity level for 0ν2β-decay of $^{100}$Mo equal to 1.24·10$^{22}$ years at confidence level of 90 % with time of measurements 90 days using one SB29 crystal.

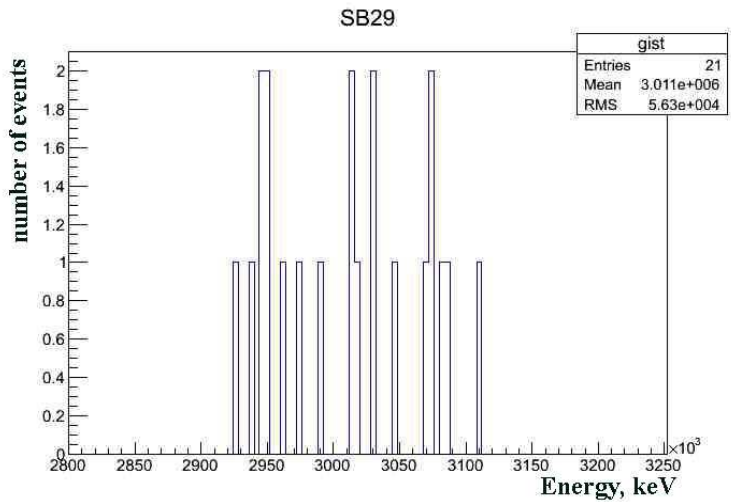

Fig. 7. Background events of SB29 crystal in the range of interest (2847-keV < E < 3223-keV) after 90 days measurements.

## 6. Conclusions

$^{40}Ca^{100}MoO_4$ scintillation elements based on enriched $^{100}$Mo and $^{40}$Ca depleted on $^{48}$Ca were produced at first time.

Internal background of scintillation elements was studied. Waveform analysis or checking average decay time of a signal can be used to solving a problem of pile-up events.

In case of S35 crystal a number of background events in the energy range of interest was decreased by 64.2 %. The S35 background index is 3.3 kev$^{-1}$ kg$^{-1}$ year$^{-1}$ (90% C.L.). It gives a new value of sensitivity of 0ν2β-decay experiment at level of $5.4 \cdot 10^{21}$ years in comparison with $4.0 \cdot 10^{21}$ years without excluding pileups [9].

The analysis of background events of SB29 crystal was done. Application of the method which was proposed in this paper allowed to select three pile-up events of total 21 in the ROI (2847-keV < E < 3223-keV). Thus, the background index of SB29 is 0.5 kev$^{-1}$ kg$^{-1}$ year$^{-1}$ (90% C.L.).

For a cryogenic experiment with 5 kg of $^{40}Ca^{100}MoO_4$ scintillation elements and 5 years of measurements the limit for 0ν2β-decay half-life is $1.8 \cdot 10^{24}$ with 90 % confidence level at the condition that the background index of the detector will be identical to one for SB29 crystal.

**Acknowledgment**

This study was supported by Federal Science and Innovations Agency of Russian Federation (Federal Aiming Program, State contract 16.523.11.3013).